\documentclass[conference]{IEEEtran}
\IEEEoverridecommandlockouts
\usepackage{cite}
\usepackage{amsmath,amssymb,amsfonts}
\usepackage{algorithmic}
\usepackage{graphicx}
\usepackage{textcomp}
\usepackage{xcolor}
\usepackage{subcaption}
\usepackage{array}
\usepackage{gensymb}
\usepackage{color, colortbl}
\usepackage[a4paper, total={7in, 8.5in}]{geometry}

\def\BibTeX{{\rm B\kern-.05em{\sc i\kern-.025em b}\kern-.08em
    T\kern-.1667em\lower.7ex\hbox{E}\kern-.125emX}}
\definecolor{Gray}{gray}{0.8}
\definecolor{LightCyan}{RGB}{0.54, 0.81, 0.94}
\definecolor{brass}{RGB}{0.71, 0.65, 0.26}

\begin{document}

\title{A Lumen Segmentation Method in Ureteroscopy Images based on a Deep Residual U-Net architecture\\
\thanks{
This work was supported by the ATLAS project. ATLAS has received funding from the European Union’s Horizon 2020 research and innovation programme under the Marie Skłodowska-Curie grant agreement No. 813782.
}
}
\author{Jorge F. Lazo$^{1,2}$, 
Aldo Marzullo$^{3}$,  
Sara Moccia$^{4,5}$, 
Michele Catellani$^{6}$,\\ 
Benoit Rosa$^{2}$, 
Michel de Mathelin$^{2}$, 
Elena De Momi$^{1}$ \\
\textit{$^{1}$ Department of Electronics, Information and Bioengineering}, Politecnico di Milano, Milan, Italy \\
\textit{$^2$ ICube, UMR 7357}, CNRS-Université de Strasbourg, Strasbourg, France \\
\textit{$^3$ Department of Mathematics and Computer Science}, University of Calabria, Rende (CS), Italy. \\
\textit{$^4$ Department of Advanced Robotics}, Istituto Italiano di Tecnologia, Genoa, Italy\\
\textit{$^5$ Department of Information Engineering}, Università Politecnica delle Marche, Ancona, Italy\\
\textit{$^6$ Istituto Europeo di Oncologia}, Milan, Italy \\
}

\maketitle

\begin{abstract}
Ureteroscopy is becoming the first surgical treatment option for the majority of urinary affections. 
This procedure is performed using an endoscope which provides the surgeon with the visual information necessary to navigate inside the urinary tract. 
Having in mind the development of surgical assistance systems, that could enhance the performance of surgeon, the task of lumen segmentation is a fundamental part since this is the visual reference which marks the path that the endoscope should follow. 
This is something that has not been analyzed in ureteroscopy data before. 
However, this task presents several challenges given the image quality and the conditions itself of ureteroscopy procedures. 
In this paper, we study the implementation of a Deep Neural Network which exploits the advantage of residual units in an architecture based on U-Net. 
For the training of these networks, we analyze the use of two different color spaces: gray-scale and RGB data images.  
We found that training on gray-scale images gives the best results obtaining mean values of Dice Score, Precision, and Recall of 0.73, 0.58, and 0.92 respectively. 
The results obtained shows that the use of residual U-Net could be a suitable model for further development for a computer-aided system for navigation and guidance through the urinary system. 
\end{abstract}

\section{Introduction}
\label{sec:introduction}

Ureteroscopy is a procedure dedicated to the exploration of the upper urinary tract to perform diagnosis and treatment of different conditions, such as kidney stones or carcinoma of the upper urinary tract. 
The procedure involves the passage of a flexible or semi-rigid ureteroscope through the urethra and bladder, and up to the ureter to the point where the stones or tumor is located~\cite{monga_2012_ureteroscopy}. 
Urinary calculi may be located anywhere in the urinary tract, but they are more common to appear in the kidneys. 
In the case of cancer, tumors can be found at any point in the renal pelvis, renal calyces, and ureters, but the most common location is the renal pelvis and renal calyces, being present in about 4-15$\%$ of the cases~\cite{latchamsetty_2006}. 
The total traveling distance from the ureter orifice to the kidneys ranges from 28cm to 34cm and the diameter of the ureter is 5mm in the pelvis~\cite{Giusti_2016}.

A review of the intraoperative complications in ureteroscopy shows that avulsion, major and minor perforation, mucosal abrasion and stricture are considered the most common ones~\cite{delaRosette_2006}. 
Minor perforation is the one with a higher frequency, appearing with an average of 1.99$\%$ of the analyzed procedures, followed by stricture (0.58$\%$), and major perforation  is the one with lower probability (0.06$\%$ of the cases).
Navigation and diagnosis inside the urinary tract are highly dependent upon the operators expertise, and image-related conditions such as the presence of image artifacts, floating debris, the low quality of the video, occlusions in the video, or image noise~\cite{ali2019endoscopy}, could add additional challenges for non-experienced operators. 
Furthermore, the heterogeneity in the anatomical structures, and the different shapes in which the lumen is deformed at different points in the urinary tract adds additional challenge to the task of segmenting the lumen. Some examples which highlight the variability in the deformation, colors and shape of the lumen are shown in Fig.~\ref{fig:sample_dataset}.

The task of lumen segmentation is a fundamental part in the development of computer vision methods for surgical assistance, since this is the reference which marks the path that the endoscope should follow. 
These assistance systems intend to deal with the mentioned challenges by enhancing relevant information to potentially increase the performance of the surgeon and minimize the probability of complications. 
Besides, the  combination  of  this visual  information  with  the information from some  other sensors such as EM sensors or Fiber Bragg Grating sensors, that  can  provide  positional  information, is  a primary  step  in  the  development  of  computer  vision  systems suitable for more advanced robot-assisted ureteroscopy~\cite{chew2020robotic}.

\newcommand{\widthfigone}{2.0cm}
\newcommand{\heighfigone}{2.0cm}

\begin{figure}[tbp]
    \begin{center}
      \begin{subfigure}[b]{0.15\textwidth}
         \includegraphics[width = \widthfigone, height =\heighfigone]{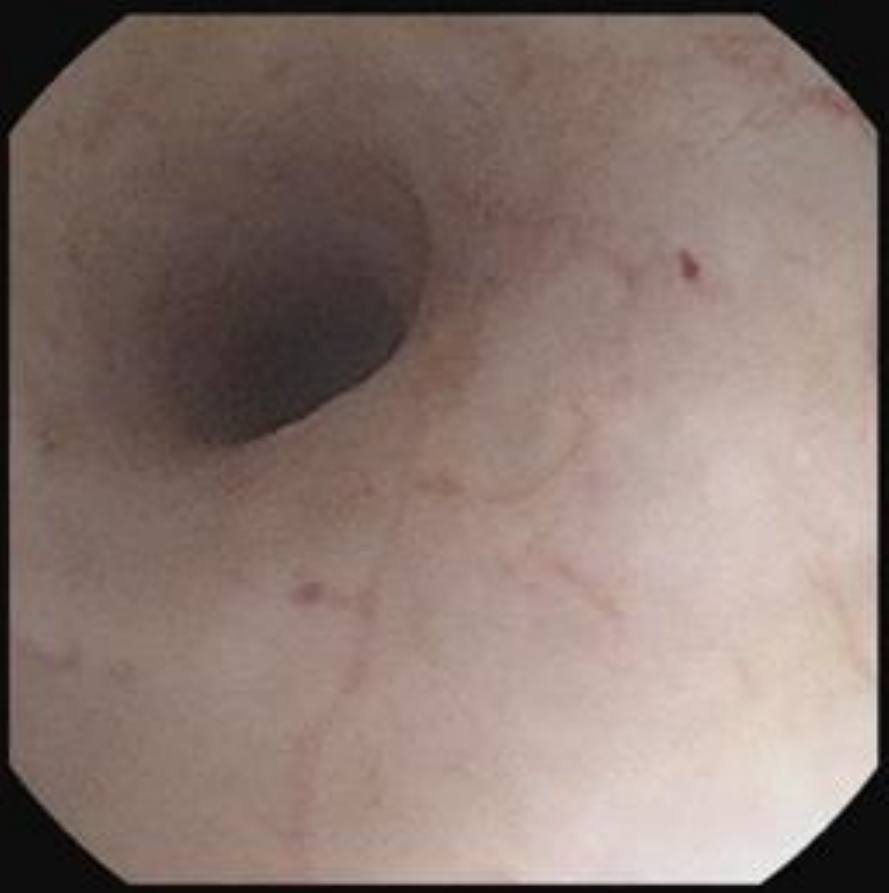} 
     \end{subfigure}
     \begin{subfigure}[b]{0.15\textwidth}
         \includegraphics[width = \widthfigone, height =\heighfigone]{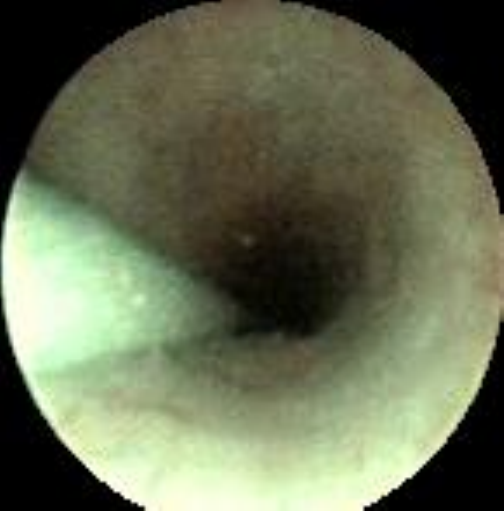} 
     \end{subfigure}
     \begin{subfigure}[b]{0.15\textwidth}
         \includegraphics[width = \widthfigone, height =\heighfigone]{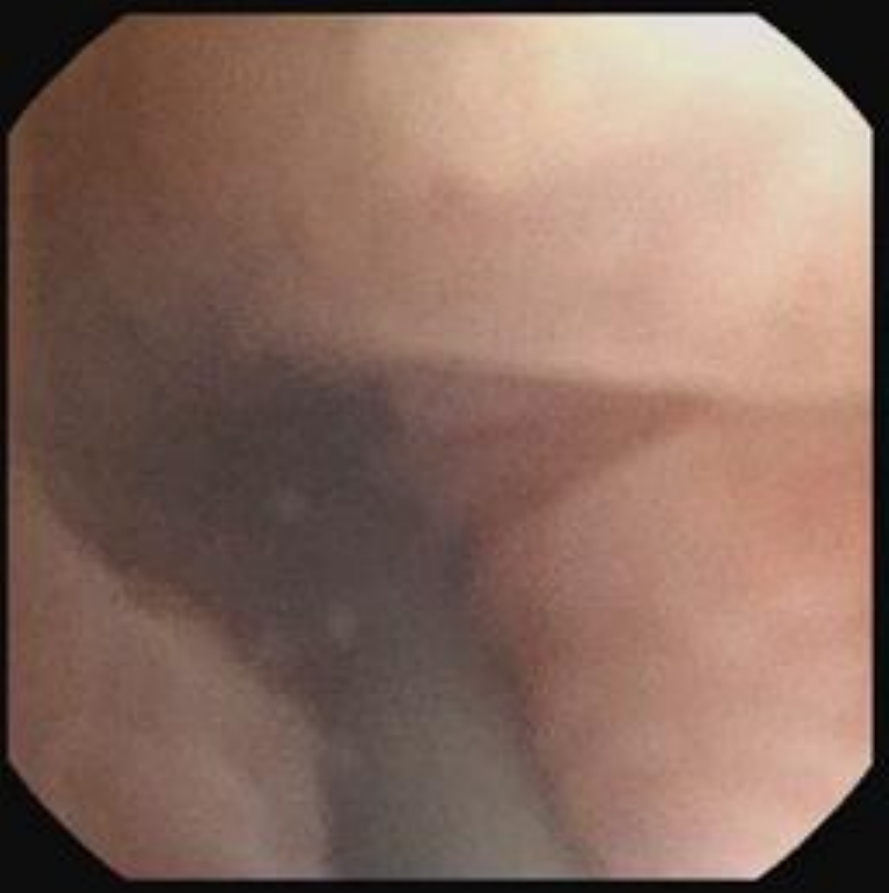} 
     \end{subfigure}
    \begin{subfigure}[b]{0.15\textwidth}
         \includegraphics[width = \widthfigone, height =\heighfigone]{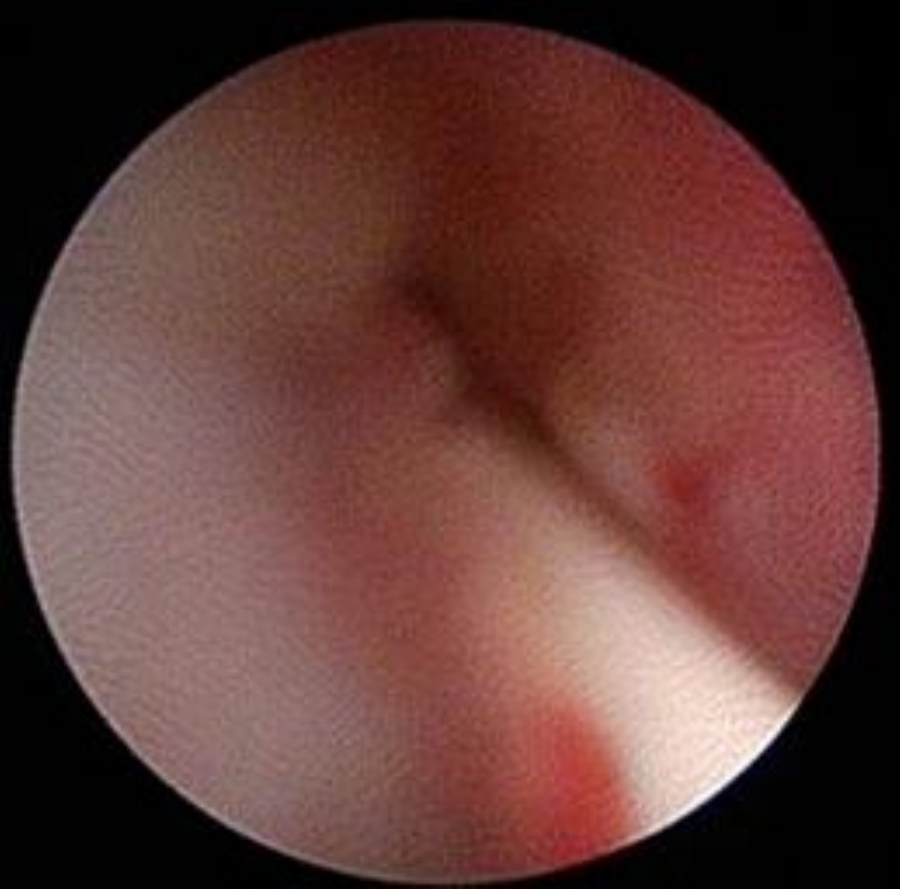} 
     \end{subfigure}
     \begin{subfigure}[b]{0.15\textwidth}
         \includegraphics[width = \widthfigone, height =\heighfigone]{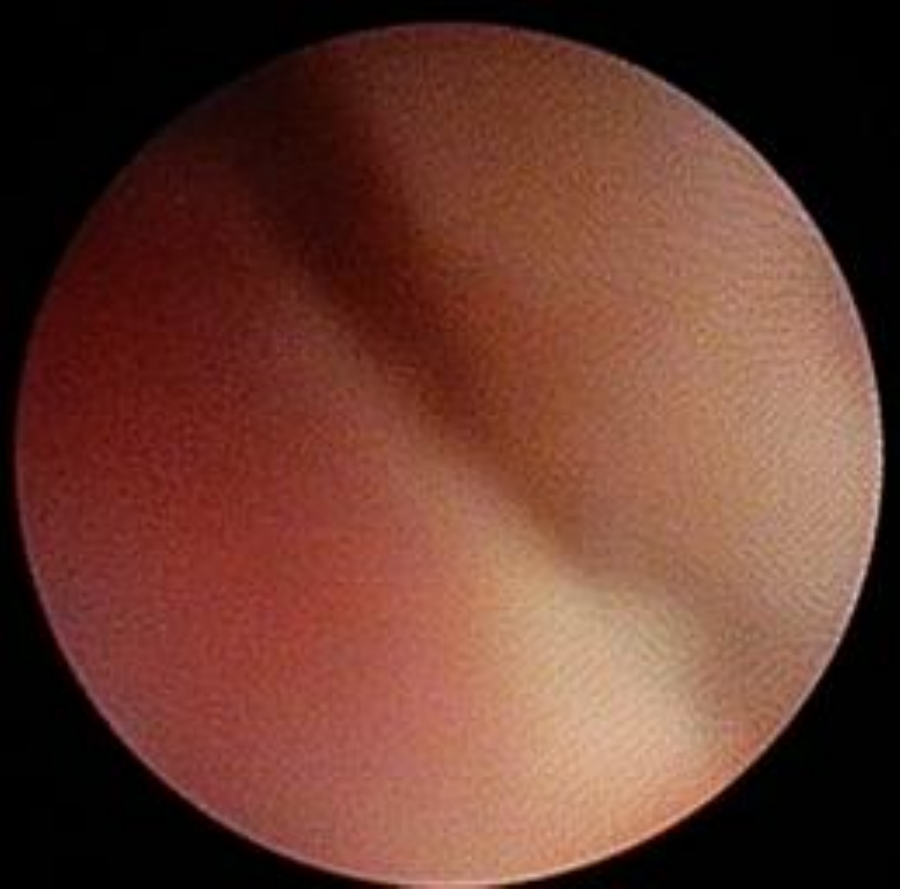} 
     \end{subfigure}
     \begin{subfigure}[b]{0.15\textwidth}
         \includegraphics[width = \widthfigone, height =\heighfigone]{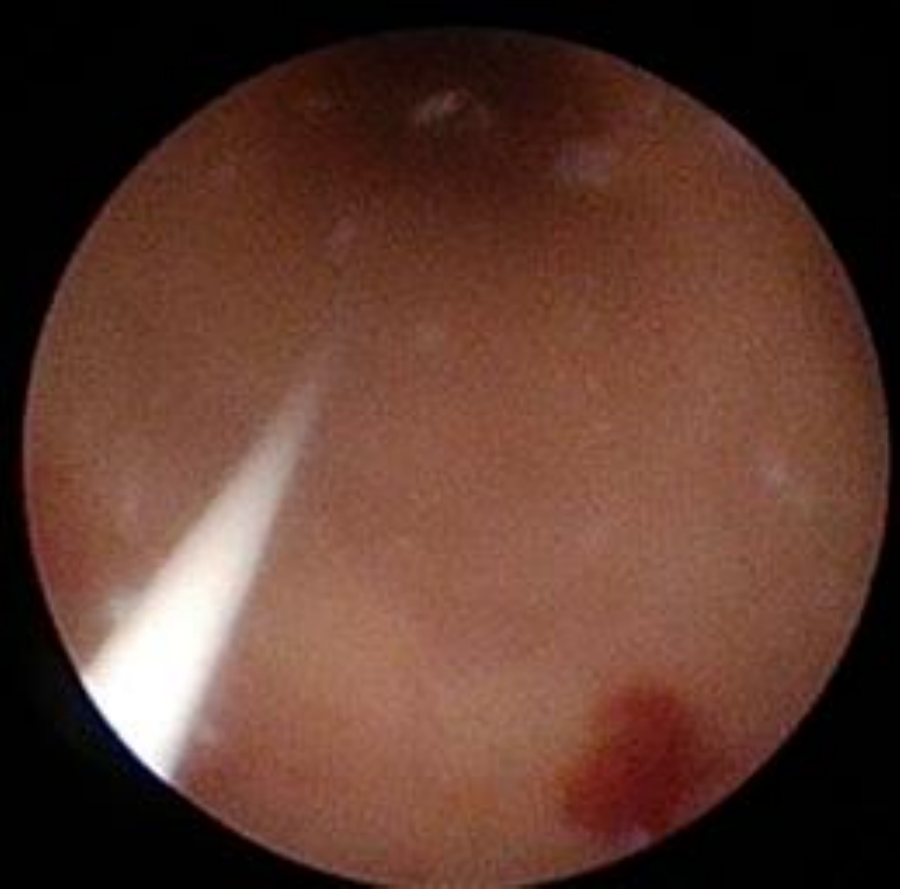}
     \end{subfigure}
    \begin{subfigure}[b]{0.15\textwidth}
         \includegraphics[width = \widthfigone, height =\heighfigone]{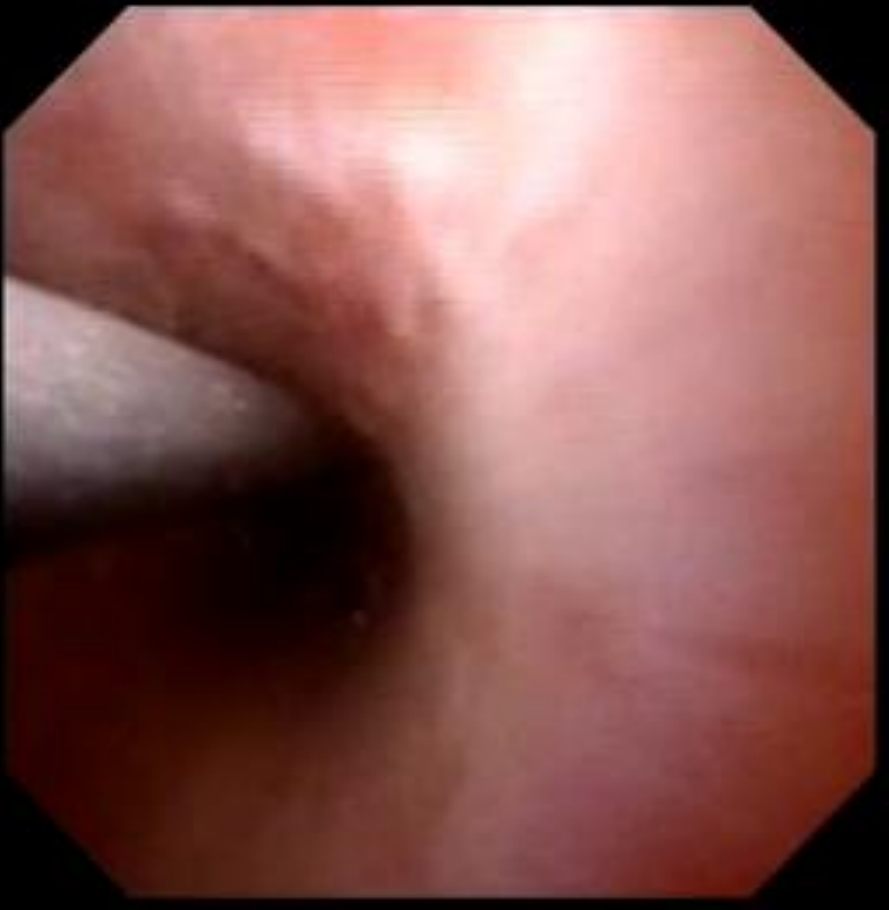}
     \end{subfigure}
     \begin{subfigure}[b]{0.15\textwidth}
         \includegraphics[width = \widthfigone, height =\heighfigone]{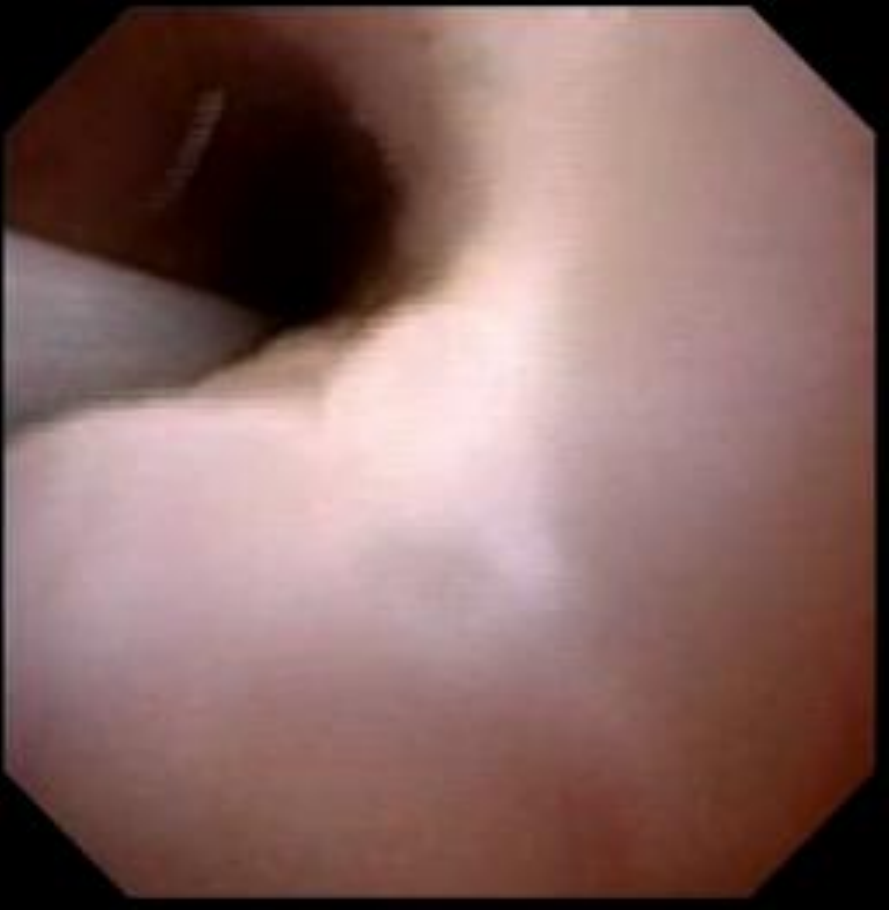} 
     \end{subfigure}
     \begin{subfigure}[b]{0.15\textwidth}
         \includegraphics[width = \widthfigone, height =\heighfigone]{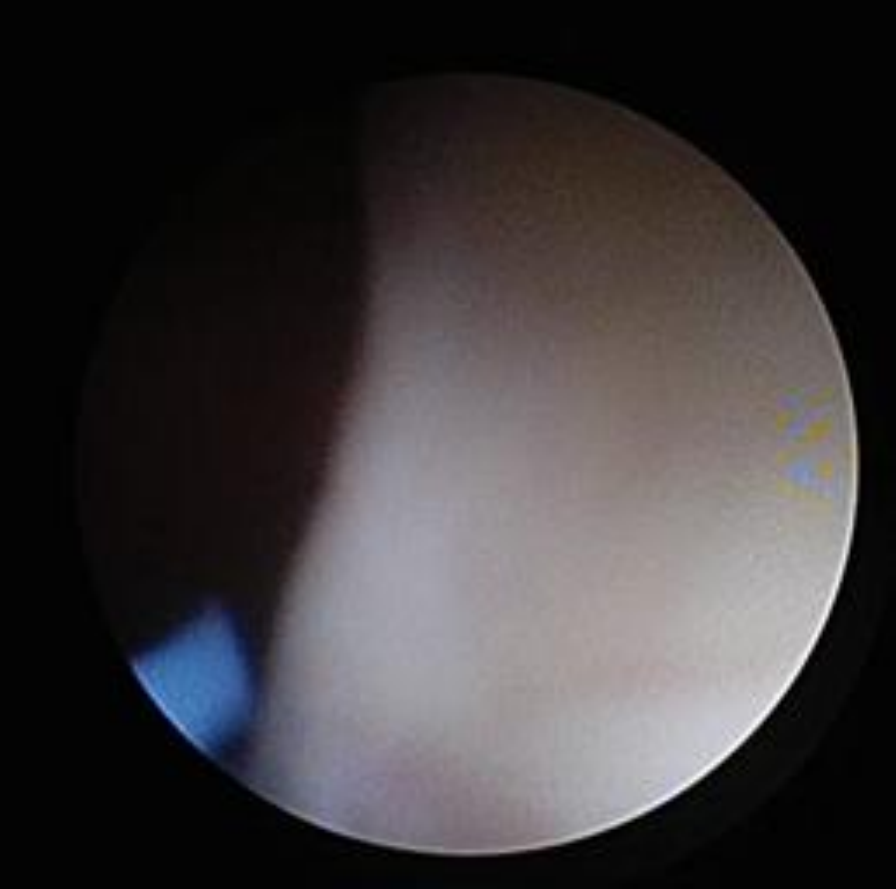} 
     \end{subfigure}
    \begin{subfigure}[b]{0.15\textwidth}
         \includegraphics[width = \widthfigone, height =\heighfigone]{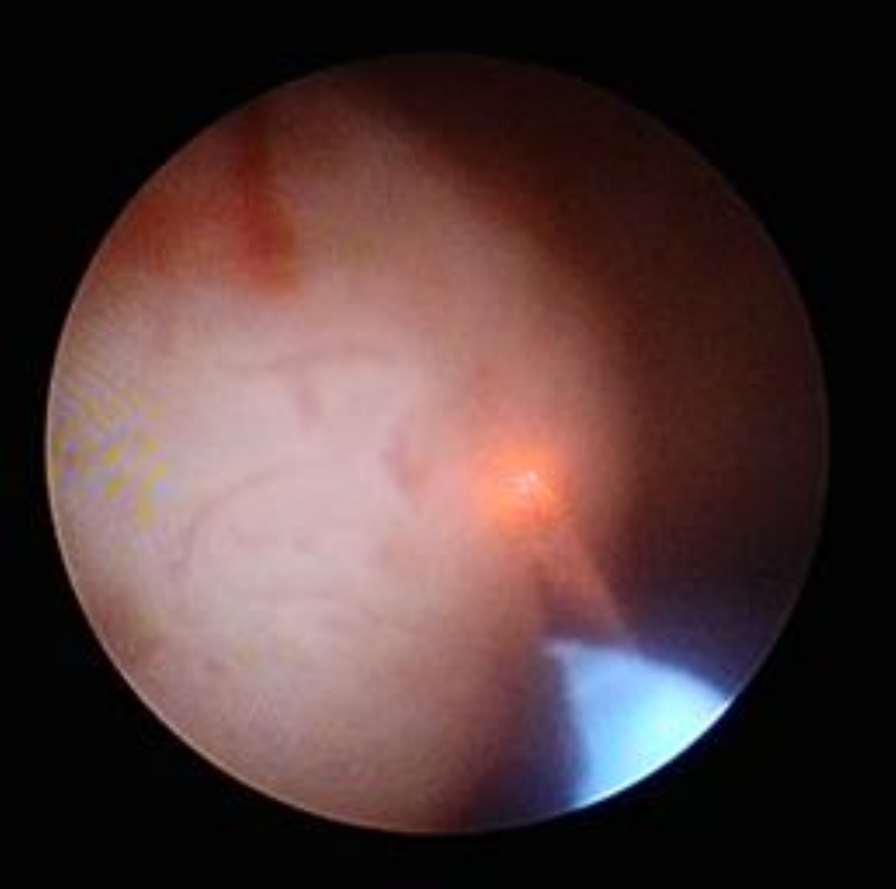} 
     \end{subfigure}
     \begin{subfigure}[b]{0.15\textwidth}
         \includegraphics[width = \widthfigone, height =\heighfigone]{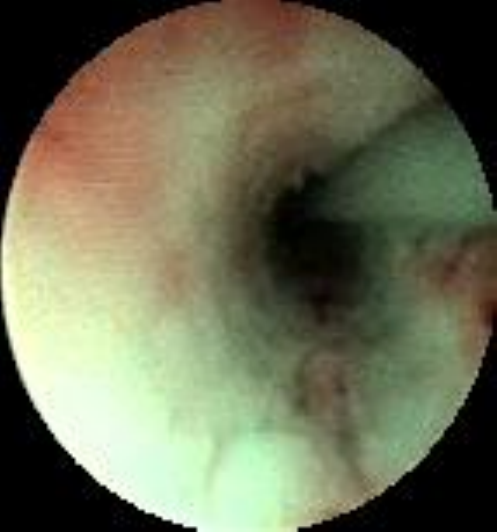} 
     \end{subfigure}
     \begin{subfigure}[b]{0.15\textwidth}
         \includegraphics[width = \widthfigone, height =\heighfigone]{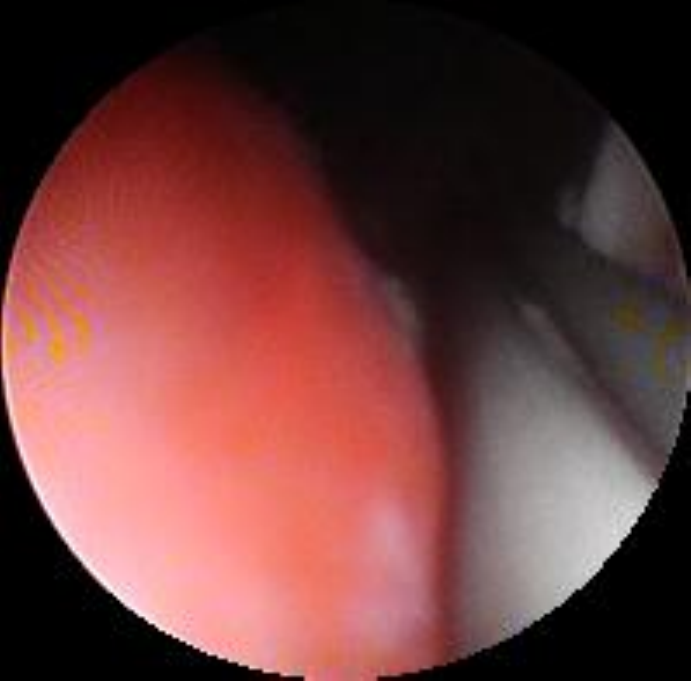}
     \end{subfigure}
    \caption{Sample of images in the dataset collected showing the variability of the hue as well as the shape and location of the tissue which sometimes merges with the tool and the guide-wire.}
    \label{fig:sample_dataset}
    \end{center}
\end{figure}

Recent advancements in the field of computer vision applied to biomedical image analysis and particularly, the use of Deep Learning (DL) methods such as Convolutional Neural Networks (CNNs) have shown remarkable results in biomedical image segmentation challenges such as  EM ISIBI 2012~\cite{arganda2015crowdsourcing}, the Longitudinal MS Lesion Segmentation Challenge~\cite{van20073d} or the Endoscopic Artefact Detection Challenge EAD2019~\cite{ali2019endoscopy}. 

In this paper, we propose the application of a variation of the well-known CNN architecture U-Net by using residual blocks~\cite{szegedy2015going}. This have shown to have beneficial effects dealing with the vanishing gradient problem~\cite{drozdzal2016importance}. 
We compare the results of training the network in the same datasets, but in different image modalities, one composed only of gray-scale images and another of RGB images to determine whether the information contained in a color-space of one single channel is enough to perform the task, or a bigger color-space is needed.  

The rest of the document is organized as follows: in Sec.~\ref{sec:related_work} we provide a survey of lumen segmentation methods used in data from other surgical procedures as well as image segmentation methods used in the analysis of ureteroscopy images, Sec.~\ref{sec:method} presents the details of the implementation used in this study, in Sec.~\ref{sec:results} the results obtained are given and discussed. Finally Sec.~\ref{sec:conclusion} concludes the paper.

\section{Related Work}
\label{sec:related_work}

Lumen segmentation has been explored in different surgical procedures such as colonoscopy procedures and also in different image modalities such as Intravascular Ultrasound images. We limit the literature research to colon procedures, which not only has a more similar anatomical structure, but the procedure itself is more similar to ureteroscopy regarding the kind of image data and instruments used during the intervention. 

The study in~\cite{gallo2012lumen} implements a intestinal detection method based on the application of a Haar-like features-based algorithm to recognize the lumen. 
A combination of the algorithm with AdaBoost to select discriminative features is  later incorporated in a cascade classifier. 
The method is sensitive to light conditions, and only works for lumen detection. 
The method proposed in~\cite{zabulis2008lumen} makes use of a coarse-to-fine version of the mean-shift algorithm to detect and track the lumen. The algorithm runs for several times with different seed points and the detected extreme points are spatially clustered. In~\cite{wang2014lumen}, two methods are presented, one based on adaptative thresholding for the detection of lumen when is not completely contracted and an alternate method for the case in which lumen is contracted. The second method is 
based on texture detection where, first adaptative thresholding is used to locate the minimum gray pixel and then the strong edges (wrinkles) of the folded intestinal wall are detected. Even if this method is capable of detecting the lumen shape no results in term of pixel accuracy are given and it is not optimal when there are considerably amounts of turbid liquid. 
The main limitation in the implementation of these methods is the fact that they are parameter-sensitive, which means that they cannot generalize very well to unseen data and in general to the inter-patient anatomy variability and imaging conditions. 
To deal with this more recent approaches have started to make use of Neural Networks-based methods.  

A benchmark for segmentation in colonoscopy images is presented in ~\cite{vazquez2017benchmark}. The authors use a multi-class image segmentation method based on the implementation of a 8-layer Fully Convolutional Network (FCN) and the classes they segment  are: background, lumen, polyp and image specularities. 
Considering the specific case of ureteroscopy images, no study related with the segmentation of lumen has been found. 
We relate this to the lack of public available datasets.
A possible reason for such fact could be difficulty of obtaining this image data and the lack of an automated or semi automated device to perform this procedure.  


The review of image segmentation methods in ureteroscopy images, even if it refers to the segmentation of different structures, is relevant for our work given the specific challenges in the analysis of uretersocopy images. Some of these challenges include a low image quality because of breathing motion~\cite{de2006renal}, presence of floating debris from the dusted tissue and the irrigation fluid used to remove it~\cite{monga_2012_ureteroscopy}. 

A kidney stone detection method using ureteroscopy images is proposed in~\cite{rosa2011algorithm} and it is based on a region growing algorithm, this algorithm requires of a seed, a similarity criterion and a stopping criterion. 
However, the definition of the criterion is not optimal since the position of the seed must be defined by the user (with the requirement that it should be placed in the center of the calculus), and the stopping criterion, which need to be optimized is empirically determined. 
A method based on Single Shot MultiBox Detector for Ureteral Orifice detection
and segmentation was proposed in~\cite{peng2019real}. 
The authors use images from resctoscopy, which have a larger field of view, to train their model and images from ureteroscope to validate it. 
Recently in~\cite{gupta2020mi} an implementation for kidney stones segmentation based on the combination of U-Net and a Deformation Vector Field which makes use of consecutive frames to determine the deformation before being fed the regular U-Net architecture has been implemented.
Nevertheless, these approaches have the peculiarity that they focus only in a specific and small part of the urinal system and they do not deal with most of the irregularities that appears in uretersocopy images.  
In our work, we deal not only with the variability of the image quality dependant on the device used in the procedure, the inter-patient anatomical variability, but also the inner-variability of the shape, texture and deformation of the ureter through the urinary tract. Furthermore, we show that the method we propose can be tested on data with the presence of elements that were not present in the data used for training, such as lasers and tumors, and still detect the lumen to some extent. 

To deal with this high-variability images we propose the use of Deep CNNs. To the best of our knowledge, CNNs have not been applied to image segmentation of the lumen in ureteroscopy images. We study the implementation of a model based in deep residual networks and compare it with other 2 models in two datasets coming from the same sources, but in different color-space. Furthermore, we implement different data augmentation techniques to improve the results obtained. 


\begin{figure}[tbp]
    \centering
    \includegraphics[width=0.5\textwidth]{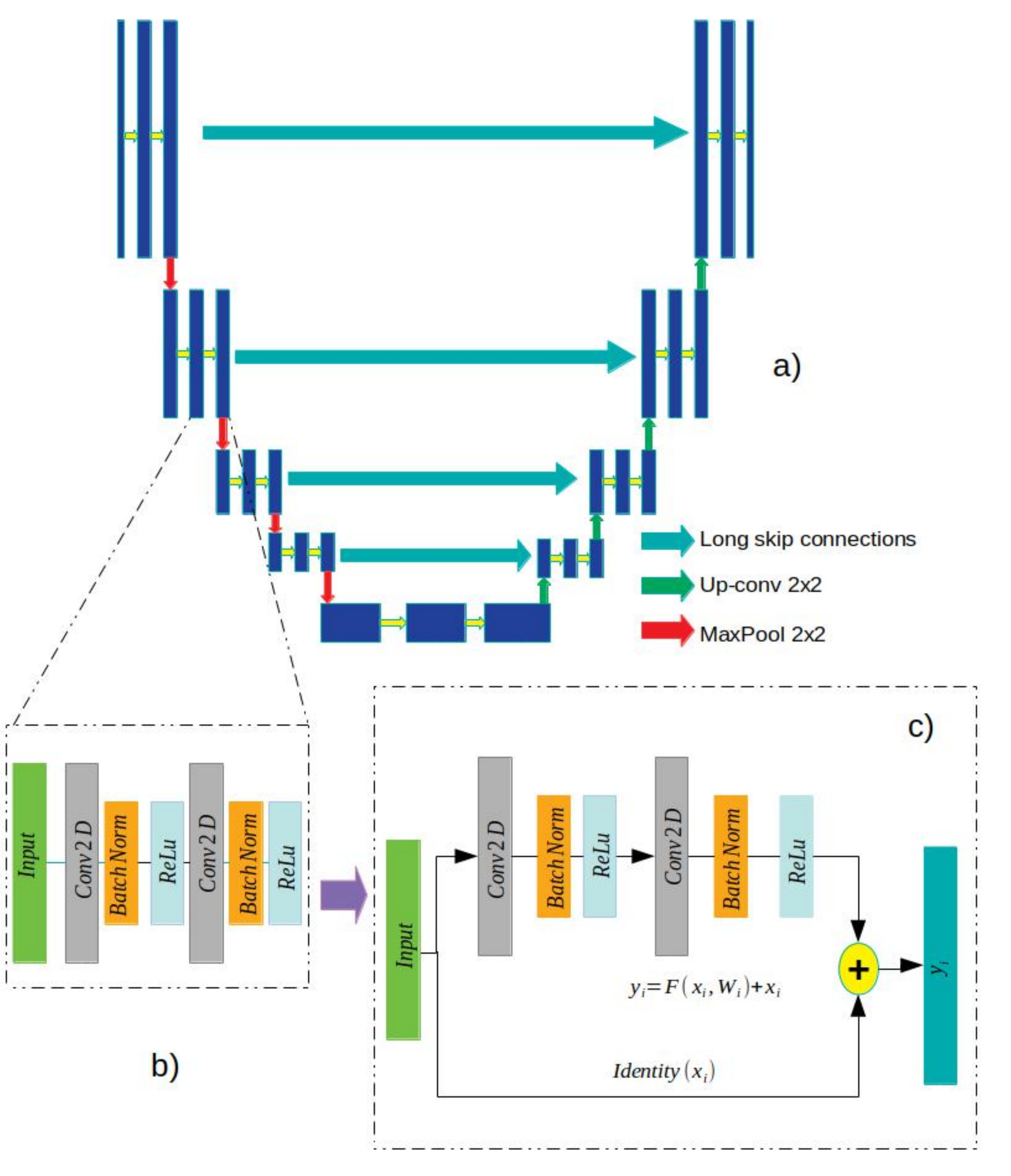}
    \caption{Comparison of the different convolutional blocks used to build the models. a) General sketch of U-Net architecture. b) Diagram of the convolutional block adding batch normalization c) residual block with short skip-connections.}
    \label{fig:unet_residual}
\end{figure}

\section{Proposed Method}
\label{sec:method}

Inspired by the paradigm set by residual networks \cite{he2016deep} which have shown to achieve state of the art performances in segmentation tasks such as  road segmentation from aerial images~\cite{zhang2018road} or nuclei segmentation~\cite{zeng2019ric}. 
We propose the implementation of a segmentation network based on U-Net architecture which makes use of residual blocks. 
Among the advantages these kind of networks presents there is the reduction in the training time, as result of the residual blocks the convergence of the learning process is faster. 
Thanks to the short-skip connections, present in this network, the spatial information propagates without degradation and in general, it has been reported that the segmentation with this architectures results in a better delineation of the borders of the figures~\cite{drozdzal2016importance}.
We compare the results obtained with the aforementioned network with two other architectures, in this case, a model similar to the standard version of U-Net, which only differs by the addition of Batch Normalization, and FCN-8 network which has been used previously for lumen segmentation. 

\newcolumntype{C}[1]{>{\centering\arraybackslash}p{#1}}

\subsection{FCN-8}
\label{subsec:FCN8}
Fully Convolutional Neural Networks (FCNs) are an extension of CNNs to deal with the task of pixel prediction. These networks make use of long skip connections in order to recover the fine-grained spatial information from the down-sampling path by merging it in the up-sampling layers. In this work we implemented a FCN8 architecture as proposed in~\cite{vazquez2017benchmark} for endoluminal scene segmentation in endoscopic images.

\subsection{U-Net}
\label{subsec:unet}
U-Net is an architecture which originally was designed for biomedical image segmentation applications~\cite{ronneberger2015u}. It is composed of a contracting path and an up-sample path which are symmetrical. The convolutional blocks of U-Net uses Convolutional Layers of kernel size 3x3 followed by 2x2 max pooling layers. The number of kernels used in each layer doubles at every block and the output of the contracting stage is connected to the feature maps in the corresponding up-sampling layer. This can ensure that the features learned in the contracting path are used in the up-sampling path for the reconstruction of the segmented images. In the implementation used in this work we added Batch Normalization layers to the output of every convolutional layer. This helps to reduce the amount of covariance shift off the hidden layers and it has the advantage that it has regularization effects, reducing overfitting and allowing the use of higher learning rates~\cite{ioffe2015batch}.

The loss function used in this implementation was based on the Dice similarity coefficient ($L_{DSC}$) defined as: 

\begin{equation}
    L_{DSC} = 1- \frac{2 TP}{2TP + FN + FP}
    \label{eq:dice_loss}
\end{equation}

where $TP$ is the number of pixels that belong to the lumen, which are correctly segmented, $FP$ is the number of pixels miss-classified as lumen, and $FN$ is the number of pixels which are classified as part of lumen but actually they are not.

\subsection{U-Net based on residual blocks}
\label{subsec:unet-residual}

Residual Blocks were first introduced in the ResNet architecture for image classification task. Their application was later expanded to image segmentation tasks and inspired the construction of networks which take the advantages of this type of convolutional blocks.
These kind of networks have proven to achieve performances above the state of the art in both, classification and segmentation tasks. 
The residual units intend to address the degradation problem by adding a skip connection between the input and the output of each convolutional block, which can be defined as: 

\begin{equation}
    y_i = h (x_i) + F (x_i, W_i)
    \label{eq:residual_output}
\end{equation}
\begin{equation}
    x_{i+1} = f(y_i)
    \label{eq:input_residual}
\end{equation}

where $x_i$ and $x_{i+1}$ are the input and the output respectively of the $i^{th}$ residual block, $F$ is the residual function, $f(y_i)$ is the activation function and $h(x_i)$ is the identity mapping function. 

The implementation used in this study consisted of using a basic residual block as defined in~\cite{drozdzal2016importance}. 
This is composed of a convolution layer followed by batch normalization and a \emph{ReLU} activation function which later was adhered to the output of the conventional convolutional block as depicted in Fig.~\ref{fig:unet_residual}. 
The loss function used in the network was the same as the used in the previous network and defined by Eq.~\ref{eq:dice_loss}.

\subsection{Performance Metrics}
\label{subsec:performance_metrics}

The performance metrics chosen were the $DSC$, the Precision ($Prec$) and Recall ($Rec$) which are defined as

\begin{equation}
    DSC = 1 - L_{DSC}
    \label{eq:DSC}
\end{equation}

\begin{equation}
    Prec = \frac{TP}{TP+FP}
    \label{eq:Prec}
\end{equation}

\begin{equation}
    Rec = \frac{TP}{TP+FN}
    \label{eq:Rec}
\end{equation}

\subsection{Dataset}
\label{subsec:dataset}

\begin{table}[tbp]
        \centering
        \caption{Detailed information about the dataset collected. The videos marked with $*$ indicate the videos that were set apart to be used only during testing.}
        \begin{tabular}{|c|C{1.5cm}|C{1.5cm}|c|}
             \hline
             & No. of annotated frames  & Image Size (pixels) & Patient No. \\ \hline
             \textbf{Video 1*}  & 7   & 356x256 & 1 \\ \hline
             \textbf{Video 2*}  & 80  & 256x266 & 1 \\ \hline
             Video 3  & 462 &  296x277 & 2 \\ \hline
             Video 4   & 245 & 256x257 & 3  \\ \hline
             Video 5   & 148 & 256x257 & 3  \\ \hline
             Video 6   & 168 & 256x257 & 3  \\ \hline
             Video 7   & 235 & 256x262  & 4 \\ \hline
             Total  & 1,445   & - & -  \\ \hline 
        \end{tabular}
        \label{tab:dataset}
    \end{table}

For this study 7 videos were collected. 
The videos were acquired from the European Institute of Oncology (IEO) at Milan, Italy.
All patients gave their informed consent for data collection and use of it for research. 
The data collection followed the ethical protocol approved by the IEO and in accordance with the Helsinky Declaration. 
The videos collected are from ureteroscopy procedures targeting upper tract tumor ablation and kidney stone removal. 
From these videos a different number of frames were extracted as described in Table~\ref{tab:dataset} and manually annotated. Some samples of the images from the dataset which depict the variability in the images is shown in Fig.~\ref{fig:sample_dataset}

A prepossessing stage included cropping of the frames to cut out the black region surrounding the field of view of the endoscopic images, and the conversion to gray-scale for the training and testing in the gray-scale dataset. The former one, was done by applying a canny filter for edge detection and subsequently an elliptical Hough transform algorithm based on \cite{xie2002new}. The final shape of the images was set to 256x256 to be consistent with the input layer of the networks used.

\begin{table}[tbp]
        \centering
        \caption{Resume of the average with its respective standard deviation of the best results obtained with the 5-fold cross validation for each of the models tested in each of the datasets: (RGB dataset white rows, grayscale-dataset gray rows).}
        \begin{tabular}{|C{0.8cm}|c|c|c|c|}
            \hline
            
            \textbf{Model} &  \textbf{$DSC$} & $Prec$ & $Rec$ & $Acc$ \\ \hline \hline
            \rowcolor{Gray} U-Net & 0.67 $\pm$ 0.01 & 0.57 $\pm$ 0.01 & 0.69 $\pm$ 0.03 & \textbf{0.94 $\pm$ 0.02}  \\ \hline
            U-Net  & 0.55 $\pm$ 0.09 & 0.43 $\pm$ 0.11 & 0.57 $\pm$ 0.07 & 0.87 $\pm$ 0.17 \\ \hline
            \rowcolor{Gray} Residual U-Net & \textbf{0.68 $\pm$ 0.05} & \textbf{0.58 $\pm$ 0.01} & \textbf{0.78 $\pm$ 0.03} & 0.92 $\pm$ 0.10\\ \hline
            Residual U-Net  & 0.59 $\pm$ 0.15 & 0.49 $\pm$ 0.03 & 0.63 $\pm$ 0.13 & 0.94 $\pm$ 0.01 \\ \hline
            \rowcolor{Gray}
            FCN-8  & 0.36 $\pm$ 0.08 & 0.25 $\pm$ 0.09 & 0.64 $\pm$ 0.01 & 0.78 $\pm$ 0.07 \\ \hline
            FCN-8  & 0.23 $\pm$ 0.04 & 0.19 $\pm$ 0.03 & 0.52 $\pm$ 0.10 & 0.64 $\pm$ 0.15 \\ \hline
        \end{tabular}
        \label{tab:results_5_fold_crossvalidations}
    \end{table}

\subsection{Training Setting}
\label{subsec:training_setting}

To handle memory limitations the networks were trained with a mini-batch size strategy. 
The networks were trained using data augmentation, the operations used for this purpose were rotation in intervals of 90$^{\circ}$, 180$^{\circ}$, and 270$^{\circ}$; horizontal and vertical flips and zooming in and out in a range of $\pm$ 0.02. 
The learning rate, and mini batch size for each of the models was chosen by trying the different combinations between several possible values of the hyper-parameters and using a 5-fold cross validation strategy with the data from patients 2, 3 and 4. 
The 5-fold process was repeated independently  with both the RGB dataset and the gray-scale dataset. The $DSC$ was set as the evaluation metric to chose the best  model with respect to the hyper-parameters. 
Adam optimization was used during all the trainings.
Once the hyper-parameter values were chosen, an additional training process was carried out using these values. 
This training was done by  using all the annotated frames obtained from patients 2, 3 and 4 for training and validation, while the data from patient 1 was used for testing. 
In total 798 frames were used for training and 417 for validation in this stage. 
The Kruskal-Wallis test on the $DSC$ was used to determine statistical significance between the different models trained.

The Networks were implemented using \textit{Tensorflow} and \textit{Keras} frameworks in Python trained on a \textit{NVIDIA Quadro M5000} GPU.

\section{Results}
\label{sec:results}

\begin{figure}[tbp]
    \centering
    \begin{subfigure}[b]{0.5\textwidth}
         \includegraphics[width = 0.99\textwidth]{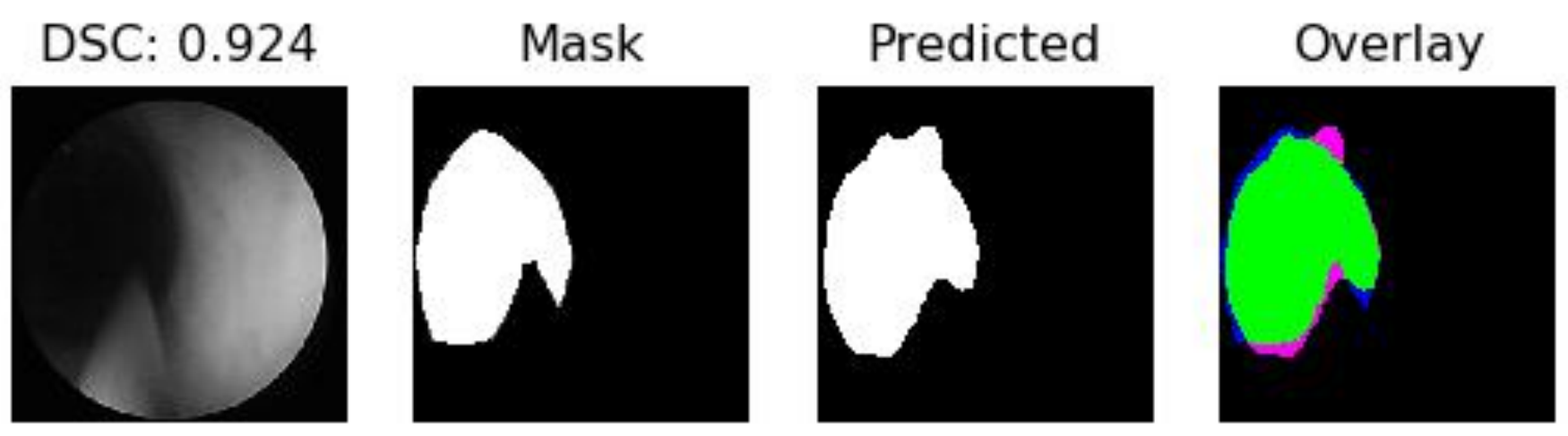} 
     \end{subfigure}
    \begin{subfigure}[b]{0.5\textwidth}
         \includegraphics[width = 0.99\textwidth]{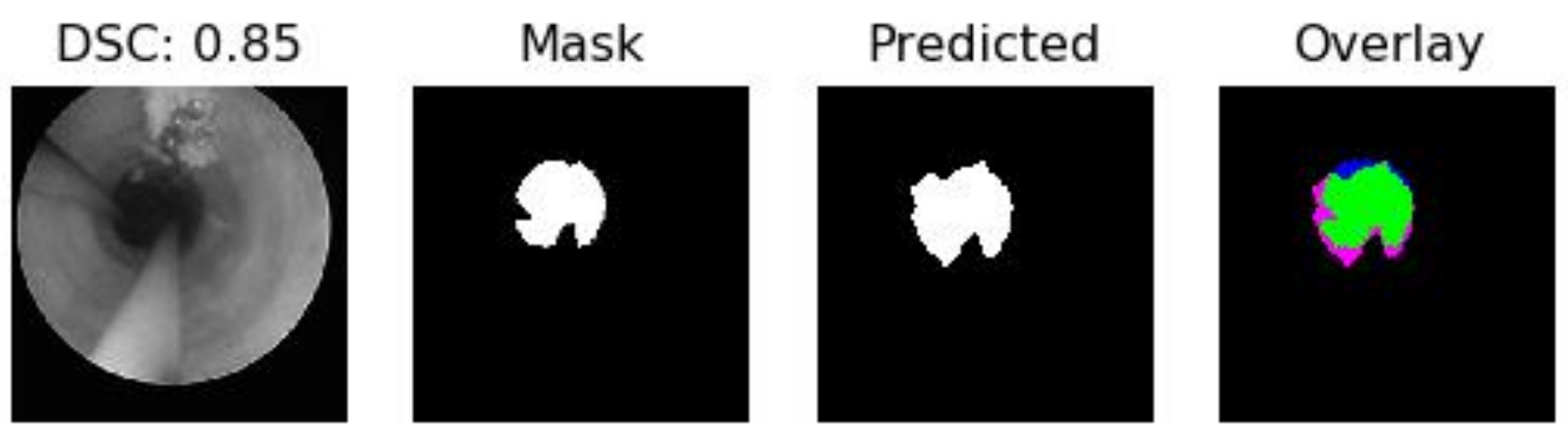} 
     \end{subfigure}
    \begin{subfigure}[b]{0.5\textwidth}
         \includegraphics[width = 0.99\textwidth]{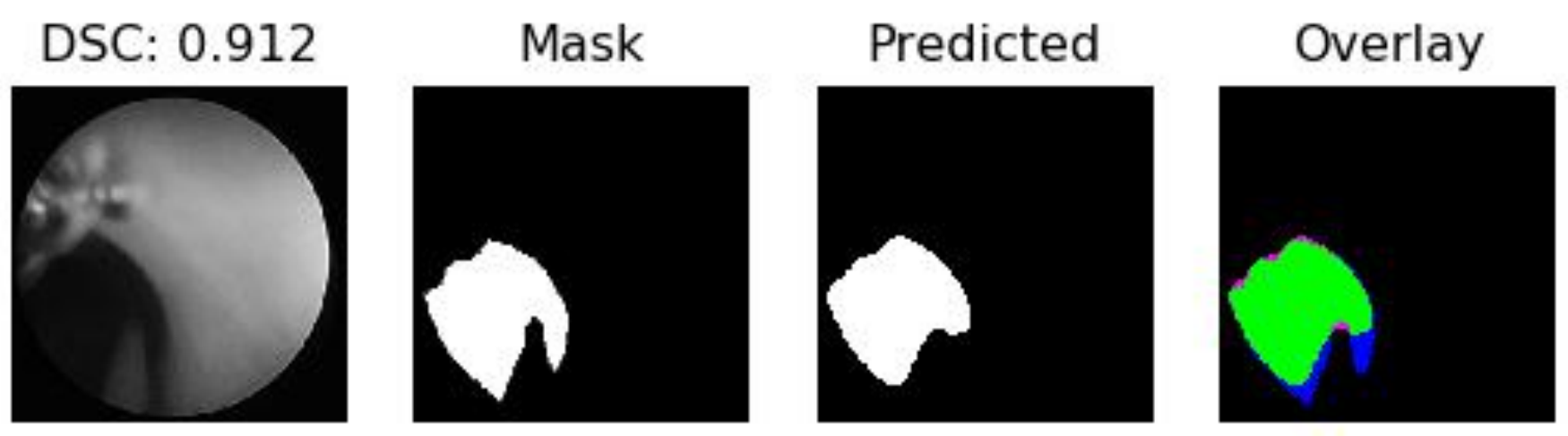} 
     \end{subfigure}
    \begin{subfigure}[b]{0.5\textwidth}
         \includegraphics[width = 0.99\textwidth]{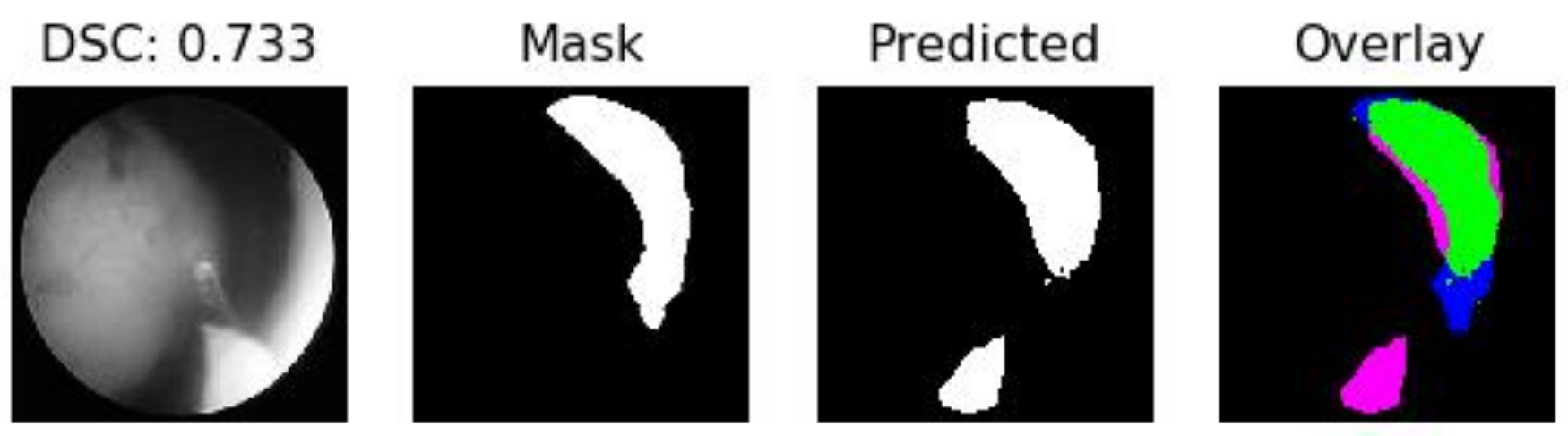} 
     \end{subfigure}
        \begin{subfigure}[b]{0.5\textwidth}
         \includegraphics[width = 0.99\textwidth]{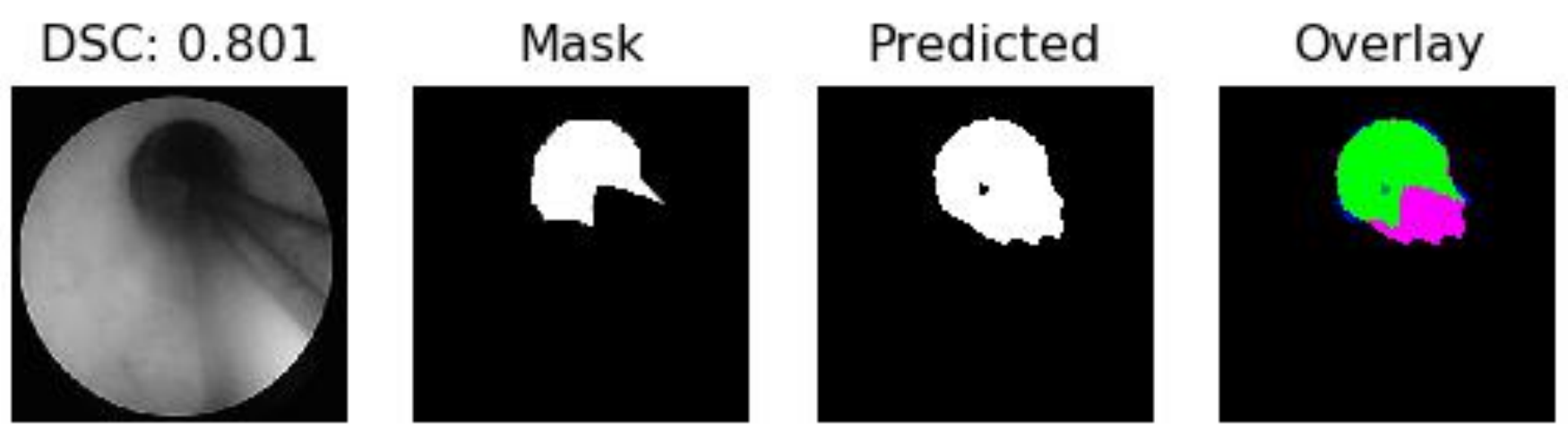} 
     \end{subfigure}
     \begin{subfigure}[b]{0.5\textwidth}
         \includegraphics[width = 0.99\textwidth]{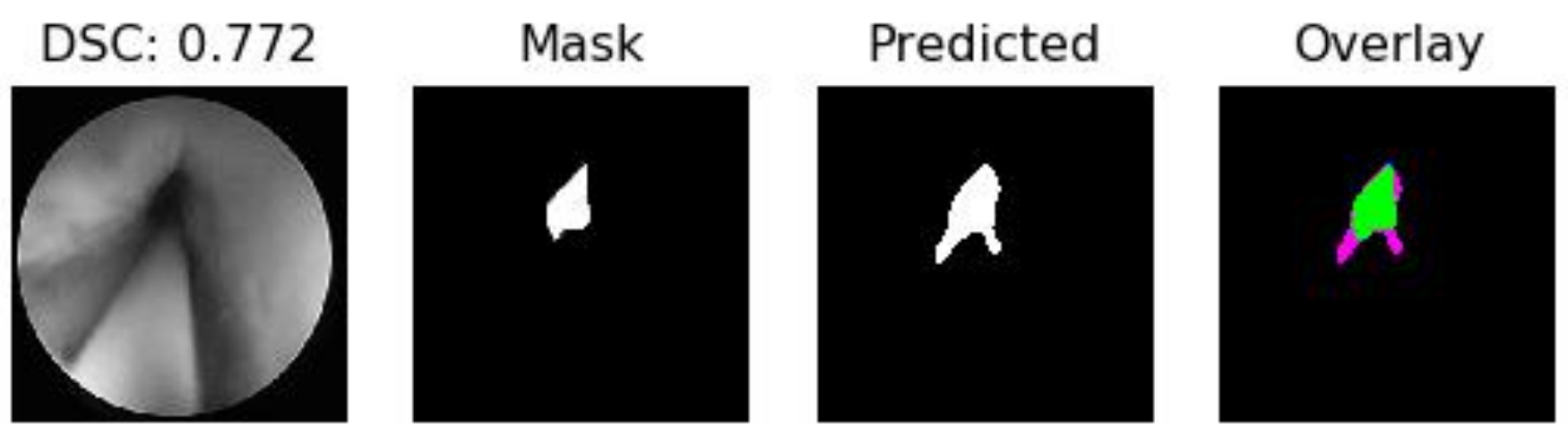} 
     \end{subfigure}
    \caption{Sample of results obtained using U-Net with skip connections with its respective dice value. The colors in the Overlay images are as follows. TP: Green, FP: Pink, FN: Blue, TN: Black.}
    \label{fig:sample_results}
\end{figure}

\begin{figure}[tbp]
    \begin{subfigure}[b]{0.5\textwidth}
        \includegraphics[width=0.99\textwidth, height=6.2cm]{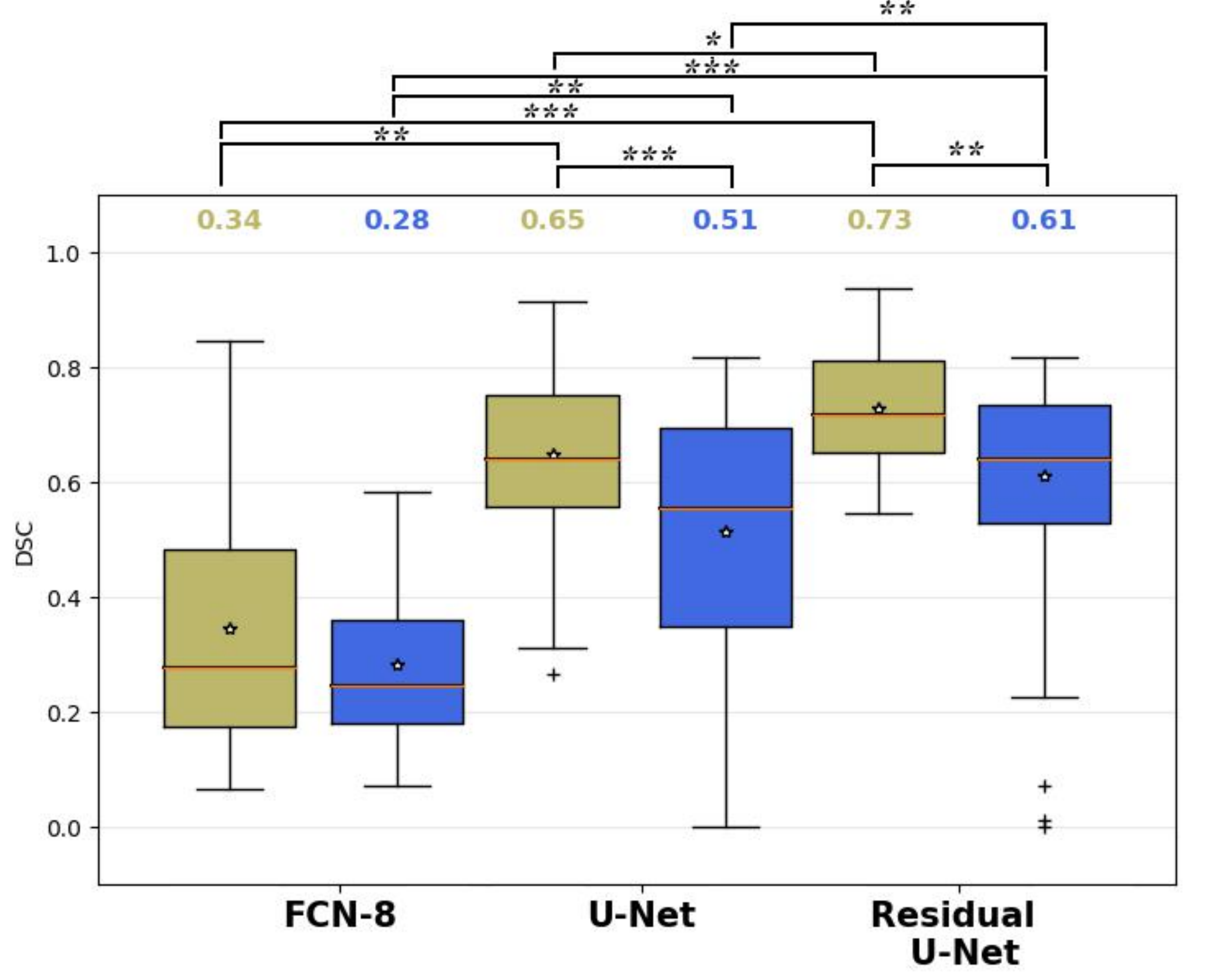}
        \caption{}
     \end{subfigure}
    \begin{subfigure}[b]{0.5\textwidth}
        \includegraphics[width=0.99\textwidth, height=5.5cm]{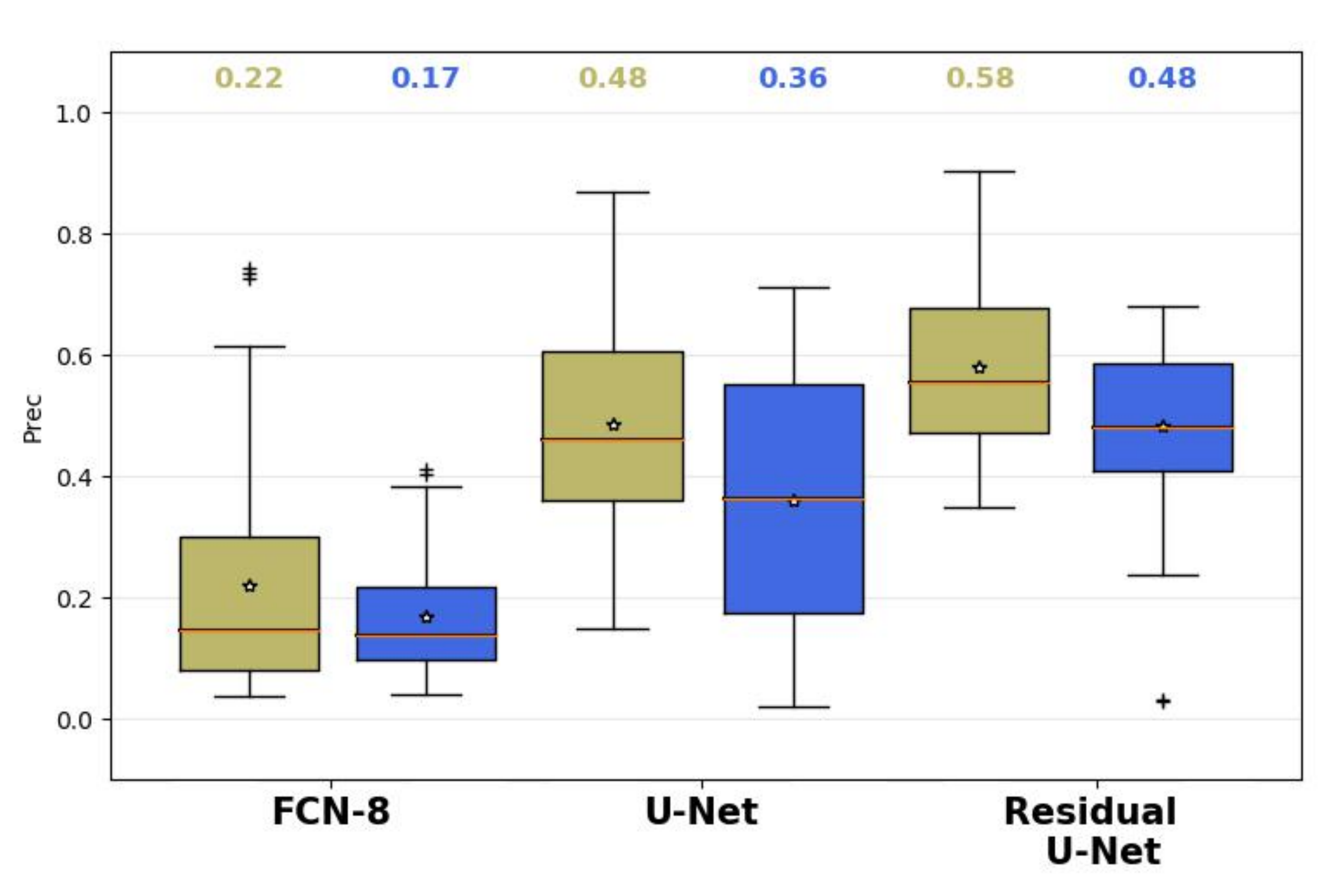}
        \caption{}
    \end{subfigure}
    \begin{subfigure}[b]{0.5\textwidth}
        \includegraphics[width=0.99\textwidth, height=6.8cm]{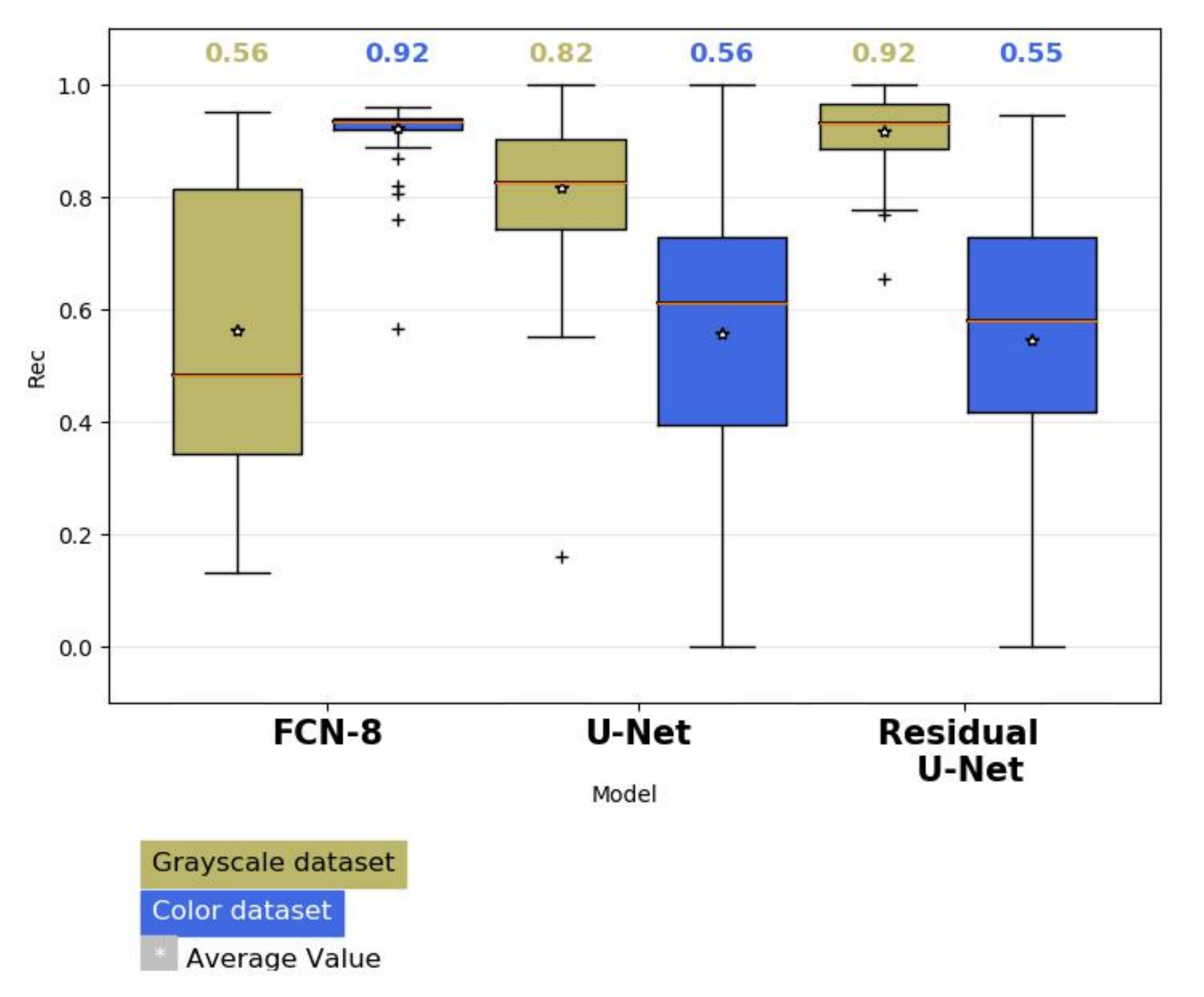}
        \caption{}
     \end{subfigure}
     \caption{\footnotesize{Box plots obtained with the different models tested. a) $DSC$, b) Precision,  c) Recall. the asterisks represent the significant difference between the different architectures in terms of the median $DSC$ Kruskal-Wallis sign-rank test (*~$p<0.05$, **~$p<0.001$, ***~$p<0.001$). The red line in the box indicates the median value.}}
    \label{fig:box_plots_results}
\end{figure}

The box plots of the $DSC$, $Prec$ and $Rec$ are depicted in Fig.~\ref{fig:box_plots_results} for each of the networks tested. From these results is possible to see that Residual U-Net has the best performance overall in $DSC$, $Prec$ and $Rec$ with average values of 0.73, 0.58, and  0.92 respectively.  
In comparison with U-Net in the gray-scale dataset Residual U-Net achieves a $DSC$ 8$\%$ better than the standard U-Net (p$<$0.05) and in the case of the RGB dataset achieves an average value 10$\%$ (p$<$0.01) better than U-Net. 
The difference between the FCN8 model and the residual U-Net results higher with a difference of 39$\%$ (p$<$0.001) for the case of the gray-scale image dataset and 33$\%$ (p$<$0.001) for the RGB one.
In general for all the 3 models it was seen that the training in gray-scale images achieve better results among the same models. In the case of the simple U-Net a difference of 14$\%$ was observed (p$<$0.001) while in the case of Residual U-Net the difference was of 12$\%$ (p$<$0.05). Even there is a difference of 6$\%$ between the use of gray-scale images and RGB ones, no statistical difference was found among them.  

\section{Discussion}
\label{sec:disc}

The collective difference between the results obtained using gray-scale images and the RGB ones might be related with the nature itself of the dataset as well as the task. 
Given that the gray-scale images might have enough information to perform the binary classification of pixels into lumen and no-lumen, the information in the RGB images provide two extra channels that does not provide any substantial new information. 
This implies the need of more parameters for the extra channels, and the addition of parameters will require a larger amount of data. 
Considering that the dataset used is not precisely large the networks  might tend to overfit even if a comprehensive exploration on the combination of hyperparameters was done, and therefore regularization techniques might be needed to overcome this problem. 

It is also worth to mention that during the training it is observed that the $DSC$ and loss curves were in general smoother for the cases in which the Residual U-Net was trained than when only the standard version of U-Net was used, and the convergence time was shorter when using the former one, just as reported in~\cite{drozdzal2016importance}.

Some sample of the results obtained with U-Net based on residual blocks in the gray-scale dataset are shown in Fig.~\ref{fig:sample_results}. 
In this case it is possible to see that at least in the majority of the cases some part of the lumen is correctly segmented and that the issues come from the border areas of the lumen where it is hard to define where it ends. 
This is in general a hard task for the networks given the fact that the lumen, which is the region of interest, has similar hue values as the region surrounding the field of view of the camera. 

Nevertheless, by using the residual-based U-net architecture it was possible to achieve a reasonable level of generalization. 
Given the fact that the video-frames used as test dataset come from a surgical-procedure of upper-tract carcinoma removal and in these, there are some elements which do not exist in the rest of the dataset such as tumors and laser beams. 
In fact, the cases in which a $DSC$ lower than 0.5 was obtained, correspond to frames in which these elements appear, or which correspond to anatomical structures which are also not present in the training dataset such as the ureteral orifice. 
This leads to encouraging results, but further explorations regarding the way in how to augment the training data to cover the same conditions present in the test dataset are needed, in case that the collection of raw data with the same conditions and elements mentioned is not feasible. 

Furthermore, the high values of recall obtained indicate that there are very low values of false negatives. 
This is important recalling the aim for which a lumen-segmentation system is intended, which is to aid the surgeon in the navigation through the urinary tract. 
Having high values of false negatives indicate that there are low possibilities that the system could lead the tip of the catheter to a position in which it could cause harm to the patient. 
Moreover, the cases in which the values of precision are not exactly high correspond mainly to the fact that the current method is missing some of the pixels in the outskirts, but the center of the lumen is always detected, which in any case, is the most important region. 

Further development to properly segment the complete area of the lumen is needed and this could be achieved by implementing more complex models which can deal with shape recognition of structures in the frames such as the one proposed in~\cite{casella2020inter} or the use of models which can exploit temporal features~\cite{colleoni2019deep}.

\section{Conclusion}
\label{sec:conclusion}
In this paper, we addressed the challenging task of lumen segmentation in the ureter, for this purpose we proposed the implementation of a Deep CNN which makes use of residual units as building blocks of a U-Net-like architecture. 
The proposed network achieves in unseen data values of $DSC$, $Prec$ and $Rec$ of 0.73, 0.58 and 0.92 respectively. 
Furthermore, the model is able, to some extent, to detect part of the lumen in images with elements, such as lasers beams and anatomical structures, such as tumors that was not present during the training, but more investigation needs to be done in order to properly generalize in such type of data.   
In conclusion the method has the potential to be further developed in order to be integrated systems which could aid surgeons in the navigation through the urinary system. 

\bibliographystyle{IEEEtran}
\bibliography{IEEEabrv,references}

\end{document}